\documentclass{article}
\usepackage{amsmath,amssymb,theorem}

\newtheorem{theorem}{Theorem}[section]   
\newtheorem{lemma}[theorem]{Lemma}
\numberwithin{equation}{section}
\numberwithin{theorem}{section}

\newcommand{\R}{\mathbb R}

\newcommand{\eps}{\epsilon}
\def\be{\begin{equation}}
\def\ee{\end{equation}}
\def\bp{\begin{pmatrix}}
\def\ep{\end{pmatrix}}
\def\bea{\begin{eqnarray}}
\def\eea{\end{eqnarray}}

\def\\{\par\medskip}

\let\0=\noindent

\newcommand{\bb}[1]{{\mathbb #1}}

\newcommand{\rmd}{\mathrm{d}}

\begin{document}

\title{Long time localization of  modified surface quasi-geostrophic equations}

\author{Guido Cavallaro, Roberto Garra,  Carlo Marchioro}

\maketitle

\date{}

\begin{abstract} 
We discuss the time evolution of a two-dimensional active
scalar flow, which extends some properties valid for a two-dimensional incompressible nonviscous
fluid. In particular we
study some characteristics of the dynamics when the field is initially
concentrated
in $N$ small disjoint regions, and we discuss the conservation in time of this
localization property. We discuss also how long this localization persists, showing that in some cases this happens for quite long times. 
\end{abstract}


\medskip\noindent
{\bf Running title:} On modified surface quasi-geostrophic equations


\bigskip

\section{Introduction and main result}
\label{sec:1}

In the present paper we discuss the long time behavior of the following Dynamical System (the so-called inviscid modified surface quasi-geostrophic equations (mSQG)):

\bigskip

\noindent let $\theta(x,t), \, x\in \R^2$ (called {\textit{active scalar}}) be the solution of the equation

\begin{equation}
\partial_t\theta +u\cdot \nabla \theta=0 \; ,
\end{equation}
where
\begin{equation}
u=(u_1,u_2)= (\partial_2 \psi,-\partial_1 \psi) 
\end{equation}
and 
\begin{equation}\label{ms0}
\theta ={\left(-\Delta\right)} ^{1-\frac{\alpha}{2}}\psi \;. 
\end{equation}

\noindent For $\alpha=0$ we have the Euler equation in dimension 2 in which $\theta$ has the meaning of vorticity, for $\alpha=1$ the Surface Quasi-Geostrophic Equation (SQG) in which $\theta$ has the meaning of temperature. The mathematical interest for the Surface Quasi-Geostrophic
Equation (SQG) started from the seminal paper of Constantin, Majda and Tabak \cite{Const1} about the singular front formation
in a model for a quasi-geostrophic flow. In this paper the authors show the formal analogy with the 3D Euler equation, studying the singular behaviour of
the solutions in the framework of the 2D SQG equation. Starting from this paper, this topic has gained a relevant interest in mathematical 
fluid mechanics, due to non-trivial problems regarding the existence and uniqueness of the solution in the inviscid case and the interesting 
dissipative generalizations. On the other hand, from the physical point of view, this model plays a relevant role in geophysical fluid 
mechanics (see \cite{Ped}), also in the studies about turbulence models (see for instance \cite{pier}).

\bigskip

\noindent The generalization of the SQG equation, namely the so-called modified SQG (mSQG) equation, corresponding to the case  $0 \leq \alpha < 1$ in \eqref{ms0}, has been recently introduced as a family of active scalar flows interpolating the 2D Euler and the SQG equations.
The time evolution of this system has been studied in many papers, establishing the existence and regularity of solutions, we refer for example to  \cite{Chae, Gancedo, Kiselev,Tan} and  references therein.
In this context, a recent topic of reseach is based on the analysis of the connection between the mSQG equation and a dynamical system
inspired by the classical point vortex model, considering the evolution of an initial datum consisting of a finite number $N$ of
strongly concentrated \textit{pseudo-vortices}, in analogy to what is done for the Euler equation
\cite{MaP93}.  We can refer to a point \textit{pseudo-vortex} system  since, as already mentioned, this dynamical
system interpolates between the Euler case (where we have point vortices) and the SQG case (where we have strongly concentrated temperature
fields). Some interesting results in this framework have been recently obtained in the literature. In \cite{CGM13,GR18, GR19, R19}, some rigorous
results about the connection between the mSQG equation and the dynamics of pseudo-vortices have been proved (the so-called localization result). Some papers have been devoted to the analysis of this dynamical system, we refer for example to \cite{ash} and \cite{BB18}.
Moreover, in \cite{Flandoli} the point vortex approximaton in the mSQG equation has been studied from the stochastic point of view. 
Finally, in \cite{Garra}, it was developed a detailed analysis of the 
growth in time of the diameter of a single patch evolving according to the modified SQG equation in the whole plane.\\

\bigskip

Let us recall that it is possible to introduce a weak form of eq.s (1.1)-(1.4):
\begin{equation}
\label{2.1}
\frac{d}{dt}\theta[f] = \theta[u \cdot \nabla f]+  \theta[\partial_t f] \; ,
\end{equation}
where  $f(x,t)$ is a bounded smooth function and
\begin{equation}
\label{2.2}
\theta [f] = \int dx \ \theta(x,t) \ f(x,t) \; .
\end{equation}
From now on we assume that the velocity $u$ decays at infinity.
In \eqref{2.1} the velocity field $u$ is given by
\begin{equation}\label{u}
u(x,t)=\int K(x-y)\theta(y,t)dy \;,
\end{equation}
where $K(x-y)= \nabla^\perp G(x-y)$, being $\nabla^\perp = (\partial_2,-\partial_1)$ and $G(x)$ the Green function of $(-\Delta)^{(1-\alpha/2)}$ with vanishing
boundary condition at infinity.

We assume that initially the active scalar is concentrated in $N$ blobs of the form
\begin{equation}
\theta_\epsilon(x,0)= \sum_{i=1}^N \theta_{i,\epsilon}(x,0)
\label{in_data}
\end{equation}
where $\theta_{i,\epsilon}(x,0)$ are functions with a definite sign such that, denoting by $\Sigma(z,r)$ the open disk of center $z$ and radius $r$,
\begin{equation}
\Lambda_{i,\epsilon}:= \text{supp} \, \theta_{i,\epsilon}(\cdot,0) \subset \Sigma(z_i,\epsilon) 
\; ; \qquad  \Sigma(z_i,\epsilon) \cap  \Sigma(z_j,\epsilon) =0 \ \ \forall i \ne j \; .
\end{equation}
Moreover we assume that
\begin{equation}
\int  dx \ \theta_\epsilon(x,0) := a_i \in \mathbb{R}
\end{equation}
independent of $\epsilon$ and 
\begin{equation}\label{mass}
|\theta_\epsilon(x,0)| \leq M \epsilon^{- 2} \; , \qquad M>0 \; , 
\end{equation}
and $u \to 0$ for ${|x| \to \infty}$.

\noindent It has been proved  in \cite{CGM13} (see also \cite{GR18}) that in general, for small $\epsilon$, the time evolution of these states has a similar form
\begin{equation}
\theta_\epsilon(x,t)= \sum_{i=1}^N \theta_{i,\epsilon}(x,t) \; ,
\label{t_data}
\end{equation}
where $\theta_{i,\epsilon}(x,t)$ are functions with definite sign such that
\begin{equation}
\Lambda_{i,\epsilon}(t):= \text{supp} \, \theta_{i,\epsilon}(\cdot, t) \subset \Sigma(z_i(t),r_t(\epsilon)) \; ,
\end{equation}
with
\begin{equation}
 \Sigma(z_i(t),r_t(\epsilon)) \cap  \Sigma(z_j(t),r_t(\epsilon))     \qquad \forall\, i \ne j \; ,
\end{equation}
being $r_t(\epsilon)$ a positive function, vanishing for  $\epsilon\rightarrow 0$, and
\begin{equation}
\dot{z}_i(t) = \sum_{j=1  , \, j\ne i}^N a_j \nabla^\perp_i G(|z_i(t)-z_j (t)|)  \; , \qquad z_i(0)=z_i \; .
\end{equation}
When all the $a_i$ have the same sign there is a global solution, while in general,
when the $a_i$ have different signs,  a collapse (that is two $z_i$ arriving at the same point) or a $z_i$ going to infinity in a finite time can happen. However in \cite{CGM13} it has been proved that these events are exceptional, and we will perform in the following the analysis
out of such zero-measure set.

\noindent We denote by {\textit{localization}} the fact that sharply concentrated initial data (of the form\eqref{in_data})  evolve in a concentrated state (of the form \eqref{t_data}). As a consequence we have a rigorous connection between the mSQG equation and a dynamical system with finite degrees of freedom (i.e. the evolution in the limit can be reduced to the evolution of $N$ concentrated fields). 

The next step in this analysis is to understand how long this localization persists.
The main aim of this work is to give a rigorous estimate of the lower bound on the first time
in which a filament reaches the boundary of a small region containing the support. Essentially, we want to generalize 
the results proved in \cite{BuM18} (for the Euler equation, i.e. $\alpha=0$) to the case of the mSQG equation ($0<\alpha<1$). This generalization is non-trivial, since the mSQG equation has a different evolution,   due to the different form of the Green function.

To each $\beta>0$ we associate the variable $T_{\eps,\beta}$ (the time horizon up to which
our localization property holds),
\begin{equation}
\label{tbe}
T_{\eps,\beta} := \min_i \sup\{t>0 \colon |x(x_0,s)-z_i(s)| < \eps^\beta\; \;\forall\, s\in [0,t]\;\; \forall\, x_0\in \Lambda_{i,\eps}(0)\}\;,
\end{equation}
being $x(x_0,s)$ the evolution of a point $x_0$ of the initial support under the dynamics (see
\eqref{tra} later).
We want to obtain a lower bound on $T_{\eps,\beta}$. 
We will confine ourselves to study the evolution in the whole plane $\bb R^2$, but the proof can be adapted to the case of a general domain. 
We can state the following:
\begin{theorem} 
\label{thm:1}
With the assumpionts above,
for each 
\begin{equation}
\beta\in \left(0,\frac{2-2\alpha}{4-\alpha}\right)
\end{equation}
there exist  $0<\epsilon_0<1$ and $\zeta_0>0$ such that
\begin{equation}
\label{tbeb1}
T_{\eps,\beta} > \zeta_0 |\log \eps | \quad\forall\,\eps\in(0,\eps_0)\;. 
\end{equation}
\end{theorem}

\medskip

\noindent Note that the upper bound for $\beta$ is less than $1/2$ (reached for $\alpha=0$).

\section{Proof of Theorem \ref{thm:1}}

The proof of the main result is based on two steps. First of all, we analyse a reduced system 
where we study the motion of a single concentrated field in an external divergence-free vector field simulating the effect of the interaction with the other blobs. Then, we will go back 
to the general case. 

\subsection{The reduced system}

Let us consider the evolution of a single blob on the whole plane according to the mSQG equation, 
simulating the interaction with the other blobs by means of an external divergence-free field
$F(x,t)$. Therefore, we have an initial datum $\theta_{\epsilon}(x,0)$ that is a function of definite sign such that 
$$\Lambda_\epsilon(0):= \text{supp} \ \theta_\epsilon(\cdot, 0) \subset \Sigma(z_* ,  \epsilon)$$
for some $z_* \in \R^2$. This initial datum evolves according to the mSQG equation with vanishing condition on the velocity at infinity ($u\rightarrow 0$ for $|x|\rightarrow \infty$).
Therefore, in this case we have that the trajectory of the fluid particle initially in 
$x_0$ is governed by
\begin{equation}\label{tra}
\frac{d}{dt}x(x_0,t)= u(x(x_0,t),t)+F(x(x_0,t),t) \; , \quad x(x_0,0)= x_0 \; .
\end{equation}  
Moreover, we can assume that the external field $F(x,t)$ is bounded and Lipschitz (with respect the space variable), i.e.,
\begin{equation}
\|F\|_{\infty}<+\infty \; , \quad |F(x,t)-F(y,t)|\leq D_t |x-y| \; , \quad D:= \sup_{t\in[0,\infty))}D_t<\infty \; .
\label{Lip_const}
\end{equation}
We also assume that the blob has intensity one and therefore, for the conservation of mass
at any time,
\begin{equation}
\theta_\epsilon(x,t)\geq 0 \; , \quad \int dx \ \theta_\epsilon(x,t) =1 \; .
\end{equation}

\smallskip

\noindent Under these assumptions, the weak form of the mSQG equation is given by
\begin{equation}\label{debole}
\frac{d}{dt}\theta[f] = \theta[(u+F)\cdot \nabla f]+  \theta[\partial_t f] \; ,
\end{equation}
and defining the center of the active scalar as
\begin{equation}
B_{\epsilon}(t)= \int x \;\theta_{\epsilon}(x,t)dx \; ,
\end{equation}
we state the following auxiliary theorem on the long time localization of a single patch, and then we go back to the general case.

\begin{theorem} 
\label{thm:2}
Let $\Lambda_\eps(t):={\rm{supp}} \,  \theta_\eps(\cdot,t)$, 
and define
\begin{equation}
\label{tbe*}
T_{\eps,\beta}^* := \sup\{t>0 \colon \Lambda_\eps(s) \subset \Sigma(B(s),  \eps^\beta) \;\; \forall\, s\in [0,t]\}\;
\end{equation}
where
\begin{equation}
\dot{B}(t) = F(B(t),t) \; , \quad B(0)= z_* \; .
\end{equation}
Then, for each 
\begin{equation}
\beta\in \left(0,\frac{2-2\alpha}{4-\alpha}\right)
\label{r_boun}
\end{equation}
 there exist $0<\eps_1<1$ and $\zeta_1>0$ such that
\begin{equation}
\label{tbeb2}
T_{\eps,\beta}^* > \zeta_1 |\log \eps | \quad\forall\,\eps\in(0,\eps_1)\;. 
\end{equation}
\end{theorem}

In order to give a complete proof of this auxiliary Theorem, we need some technical lemmas.
The proof of the Theorem will be given at the end of the subsection.

\noindent First of all we recall the known form of the Green function in the mSQG equation.
The Green function of $(-\Delta)^{(1-\alpha/2)}$ in $\R^2$ with vanishing boundary conditions at infinity is:
\begin{equation}
\label{2.8}
G(r)= \varphi (\alpha) r^{-\alpha} \; , \quad  \varphi(\alpha) = \frac{1}{\pi 2^{(2-\alpha)}}\frac{\Gamma(\alpha/2)}{\Gamma(\frac{2-\alpha}{2})} \; , \quad r=\sqrt{x^2_1+x^2_2} \; ,
\end{equation}
where $\Gamma(\cdot)$ denotes the Euler Gamma function and $\alpha \in (0,1)$.

Now, we give the estimates on the evolution of the center of the active scalar $B_\epsilon(t)$ and
the moment of inertia with respect to $B_\epsilon(t)$ defined as
\begin{equation}
I_\epsilon(t) =\int dx |x-B_\epsilon (t)|^2 \, \theta_\epsilon(x,t) \; .
\end{equation}

\begin{lemma}
For any $t\geq 0 $, we have 
\begin{equation}
I_\epsilon(t)\leq 4 \epsilon^2 \exp\bigg[2\int_0^t ds \ D_s\bigg]
\end{equation}
and 
\begin{equation}
|B_\epsilon(t)-B(t)|\leq 2 \epsilon\left(1+\int_0^t ds \ D_s\right) \exp\bigg[\int_0^t ds \ D_s\bigg] \; .
\end{equation}
\end{lemma}

We neglect the proof of this lemma because it is immediate to see that it coincides to the Euler case proved in \cite{BuM18} (Lemma 2.3). We observe that if $F$ is null,
$B_\epsilon(t)$ and $I_\epsilon(t)$ are conserved along the motion in the mSQG case as well
as in the Euler case.\\
We now introduce the positive parameter $\gamma$ and study the system for $t\in[0,\gamma|\log\epsilon|]$. From  the definition of $D$ \eqref{Lip_const}, we have the following estimates for $t\in[0,\gamma|\log\epsilon|]$ and $\gamma$ small enough:
\begin{equation}
I_\epsilon\leq 4 \epsilon^\delta \; ,
\label{I_eps}
\end{equation}
\begin{equation}
|B_\epsilon(t)-B(t)|\leq 2(1+D\gamma|\log\epsilon|)\epsilon^{\delta/2} \; ,
\end{equation}
with $\delta = 2-2D\gamma>0$.
These bounds imply that for fixed small $\epsilon$ the main part of the scalar field remains
concentrated around the center of the active scalar. We now study the growth in time of the distance
from the center of the active scalar, in order to prove that all the filaments
of the scalar field remain close to $B_\epsilon(t)$.

The next two lemmas request more attention and essentially differ from the Euler case, due
to the different form of the kernel determining the evolution of the velocity field in 
the mSQG equation.

{\textit{A warning on the notation}}. Hereafter  in the paper
we denote by $C$
a generic positive constant (eventually changing from line to line) which is independent of the parameter $\epsilon$
and the time $t$, and by $C_i$ ($i$ an integer index) a positive constant as before, which needs to be quoted elsewhere.

\begin{lemma}
Let be $R_t$ defined as follows
\begin{equation}\label{Rdit}
R_t := \max\{|x-B_\epsilon(t)|:x\in \Lambda_\epsilon(t)\} \; ,
\end{equation}
where $\Lambda_\epsilon (t)= \text{supp} \ \theta_\epsilon(\cdot,t)$.
Assume that $x_0\in \Lambda_\epsilon(0)$ is such that
\begin{equation}
|x(x_0,t)-B_\epsilon(t)|=R_t \; .
\label{R_ti}
\end{equation}
Then 
\begin{equation}
\frac{d}{dt}|x(x_0,t)-B_\epsilon(t)|\leq 2 D_t R_t +\frac{C_1 \ I_\epsilon(t)}{R_t^{3+\alpha}}+
\frac{C_2}{1-\alpha}\left[\left(M\epsilon^{-2}\right)^{\frac{1+\alpha}{1-\alpha}} \, m_t(R_t/2)\right]^{\frac{1-\alpha}{2}}
\label{eq_maxspost}
\end{equation}
where 
\begin{equation} 
m_t(h) =\int_{|y-B_\epsilon(t)|>h}\ dy \ \theta_\epsilon(y,t) \; .
\label{massa_bordo}
\end{equation}
\end{lemma}

\textit{Proof.} First of all, observing that 
\begin{equation}
 \dot{B}_\epsilon(t) = \int dx F(x,t) \, \theta_\epsilon (x,t) \; ,
\label{Bdot}
\end{equation}
by using \eqref{tra} and the assumption of unitary intensity, we have 
\begin{align}
&\frac{d}{dt}|x(x_0,t)-B_\epsilon(t)|=\left(u(x,t)+F(x,t)-\frac{d}{dt}B_\epsilon(t)\right)
\cdot\frac{x-B_\epsilon(t)}{|x-B_\epsilon(t)|}\\
\nonumber &= \frac{x-B_\epsilon(t)}{|x-B_\epsilon(t)|}\cdot\int dy \ [F(x,t)-F(y,t)]\theta_\epsilon(y,t)\\
&\,\,\,\,\,\,+\frac{x-B_\epsilon(t)}{|x-B_\epsilon(t)|}\cdot \int dy K(x-y)
\theta_\epsilon(y,t) \; . 
\end{align}
The first term is easily bounded by the Lipschitz property of the external field
$F(x,t)$ and \eqref{Rdit}; we have 
\begin{equation}
\bigg|\int \ dy [F(x,t)-F(y,t)]\theta_\epsilon(y,t)\bigg|\leq D_t\int dy |x-y|\theta_\epsilon(y,t)
\leq 2 D_t R_t \; .
\end{equation}
For the second term, we split the integration domain into two parts:
\begin{equation*}
A_1=\Sigma(B_\epsilon(t),  R_t/2) \; ,  \qquad A_2=\Sigma(B_\epsilon(t),  R_t))\setminus \Sigma(B_\epsilon(t),  R_t/2) \; .
\end{equation*}
Therefore, we are going to study in detail the two integrals
\begin{equation}
H_1 = \frac{x-B_\epsilon(t)}{|x-B_\epsilon(t)|}\cdot \int_{A_1} dy \ K(x-y)
\theta_\epsilon(y,t)
\label{acca_uno}
\end{equation}
over the disk $A_1$ and
\begin{equation}
H_2 = \frac{x-B_\epsilon(t)}{|x-B_\epsilon(t)|}\cdot \int_{A_2} dy  \ K(x-y)
\theta_\epsilon(y,t)
\end{equation}
over the external annulus $A_2$, that is the region near the boundary of the support.
Here we have the main technical difference from the Euler case, due to the 
different kernel in the evolution of the velocity field.\\
\noindent Let us consider the contribution related to $A_1$.\\
\noindent Introducing the variables $x' = x-B_\epsilon(t)$, $y' = y-B_\epsilon(t)$, using Lemma 2.1 and the fact that
$x' \cdot (x'-y')^{\perp}= -x'\cdot y'^{\perp}$, we obtain 
\begin{equation}
H_1 = \alpha\varphi(\alpha)\int_{|y'|\leq R_t/2} \ dy' \frac{x'\cdot y'^\perp}{|x'||x'-y'|^{2+\alpha}}
\theta_\epsilon(y'+B_\epsilon (t)) \; .
\label{aa}
\end{equation}
Since
\begin{equation*}
\int \ dy' y'^\perp \theta_\epsilon(y'+B_\epsilon(t))= 0 \; ,  
\end{equation*}
by adding and subtracting 
$$
\frac{x'\cdot y'^\perp}{|x'|} \frac{1}{|x'|^{\alpha+2}} \, \theta_\epsilon(y'+B_\epsilon (t))
$$
in the integrand of (\ref{aa}),
we can write
\begin{equation}
H_1 = H_1'-H_1'' \; ,
\end{equation}
where
\begin{align}
\nonumber & H_1' = \alpha\varphi(\alpha)\int_{|y'|\leq R_t/2} dy' \frac{x'\cdot y'^\perp}{|x'|}\frac{|x'|^{\alpha+2}-|x'-y'|^{\alpha+2}}{|x'-y'|^{\alpha+2} |x'|^{\alpha+2}}\theta_\epsilon(y'+B_\epsilon(t))\\
\nonumber & H_1'' = \alpha\varphi(\alpha)\int_{|y'|> R_t/2} dy' \frac{x'\cdot y'^\perp}{|x'|^{3+\alpha}}\theta_\epsilon(y'+B_\epsilon(t)) \; .
\end{align}
By using the Lagrange theorem to the function $|z|^{\alpha+2}$ we have
$$
\left| \frac{x'\cdot y'^\perp}{|x'|}\frac{|x'|^{\alpha+2}-|x'-y'|^{\alpha+2}}{|x'-y'|^{\alpha+2} |x'|^{\alpha+2}} \right|
\leq  \left| \frac{x'\cdot y'^\perp}{|x'|}\frac{ C |\xi|^{\alpha+1} \, |y'|}{|x'-y'|^{\alpha+2} |x'|^{\alpha+2}} \right|
$$
where $\xi$ is a suitable point on the segment joining $x'$ and $x'-y'$.  

\noindent From \eqref{Rdit}, $|x'| = R_t$, $|y'|\leq R_t/2$ and therefore 
$|x'-y'|\geq R_t/2$.
Thus, we have 
\begin{equation}
|H_1'| \leq \frac{C}{ R_t^{\alpha+3}}\int_{|y'|\leq R_t/2}dy' |y'|^2 \theta_\epsilon(y'+B_\epsilon(t))\leq \frac{C\, I_\epsilon(t)} {R_t^{\alpha+3}} \; .
\label{acca_uno'}
\end{equation}
Regarding the integral $H_1''$, that is restricted to $|y'|\geq R_t/2$ and also to $|y'|\leq R_t$
(by \ref{Rdit}) we have 
\begin{equation}
|H_1''|\leq \frac{C \, I_\epsilon(t)}{R_t^{\alpha+3}} \; ,
\label{acca_uno''}
\end{equation}
where we used the Chebyshev inequality.\\
Going back to the integral $H_2$, we get
\begin{equation}
|H_2|\leq \alpha\varphi(\alpha)\int_{A_2} dy \frac{1}{|x-y|^{\alpha+1}}\theta_\epsilon(y,t) \; .
\end{equation}
The maximum of the last integral is obtained by rearranging the active scalar's mass 
as close as possible to the singularity.
By using the assumption \eqref{mass} we have 
\begin{equation}
|H_2| \leq\alpha\varphi(\alpha) \, M\epsilon^{-2}\int_{\Sigma(0, r)}dy\frac{1}{|y|^{\alpha+1}} = 
2\pi \alpha\varphi(\alpha) \, M\epsilon^{-2}\frac{r^{1-\alpha}}{1-\alpha} \; ,
\end{equation}
where the radius $r$ is such that $M\epsilon^{-2} \pi r^2 = m_t(R_t/2)$.\\
\noindent Taking all these estimates together we finally obtain the claimed result
(note that $\lim_{\alpha\to 0} [\alpha\varphi(\alpha)]=\frac{1}{2\pi}$). $\square$
 
 \medskip
 
Let us define
\begin{equation}
\bar\beta = \frac{\delta\left(1-\frac{\alpha}{2}\right)-\alpha}{4-\alpha} \;,
\label{barbeta}
\end{equation}
where $\delta = 2-2D\gamma>0$. By choosing $\gamma$ suitably small we can obtain $\bar\beta>0$
(in addition it is less than $1/2$). We state the following lemma.
\begin{lemma}
\label{lem:3}
Let $m_t$ be defined as in \eqref{massa_bordo}. For each $\beta\in (0,\bar\beta)$  and $\ell>0$ there exists $\gamma>0$ such that
\begin{equation}
\label{smt}
\lim_{\eps\to 0} \eps^{-\ell} m_t(\eps^\beta) = 0 \quad \forall\, t \in [0,\gamma|\log\eps|]\;. 
\end{equation}
\end{lemma}

\textit{Proof}.
Given $h>0$, let $x\mapsto W_h(x)$, $x\in \bb R^2$, be a nonnegative smooth function,  depending only on $|x|$, such that
\begin{equation}
\label{W1}
W_h(x) = \begin{cases} 1 & \text{if $|x|\le h$,} \\ 0 & \text{if $|x|\ge 2h$}, \end{cases}
\end{equation}
and, for some $C_3>0$,
\begin{equation}
\label{W2}
|\nabla W_h(x)| < \frac{C_3}{h}\;,
\end{equation}
\begin{equation}
\label{W3}
|\nabla W_h(x)-\nabla W_h(x')| < \frac{C_3}{h^2}\,|x-x'|\;. 
\end{equation}

We define the quantity
\begin{equation}
\label{mass 1}
\mu_t(h) = 1 - \int\! \rmd x \, W_h(x-B_\eps(t))\, \theta_\eps (x,t)\;,
\end{equation}
which is a mollified version of $m_t$, satisfying
\begin{equation}
\label{2mass 3}
\mu_t(h) \le m_t(h) \le \mu_t(h/2)\;.
\end{equation}
In particular, it is enough to prove \eqref{smt} with $\mu_t$ instead of $m_t$. 

To this purpose, we study the time derivative of $\mu_t(h)$. By applying \eqref{debole} with test function $f(x,t) = W_h(x-B_\eps(t))$, and recalling 
$$
u(x,t) = \int\!\rmd y\, K(x-y)\theta_\eps(y,t)
$$ 
and \eqref{Bdot}, we have, 
\begin{equation}
\label{mass 4}
\begin{split}
\frac{\rmd}{\rmd t} \mu_t(h) & = - \int\! \rmd x\, \nabla W_h(x-B_\eps(t)) \cdot [u(x,t)+ F(x,t) - \dot B_\eps(t)]\,\theta_\eps(x,t) \\ & =  - H_3 - H_4\;, 
\end{split}
\end{equation}
with
\begin{equation*}
\begin{split}
H_3 & = \int\! \rmd x\, \nabla W_h(x-B_\eps(t)) \cdot \int\!\rmd y \, K(x-y)\, \theta_\eps(y,t)\, \theta_\eps(x,t) \\ & = \frac 12 \int\! \rmd x \! \int\! \rmd y\, \theta_\eps(x,t)\,  \theta_\eps(y,t) \, [\nabla W_h(x-B_\eps(t))-\nabla W_h(y-B_\eps(t))] \cdot K(x-y) \;, \\ H_4 & = \int\! \rmd x\, \nabla W_h(x-B_\eps(t)) \cdot \int\!\rmd y \,[F(x,t)-F(y,t)]\, \theta_\eps(y,t)\, \theta_\eps(x,t)\;,
\end{split}
\end{equation*}
where the second expression of $H_3$ is due to the antisymmetry of $K$.

Concerning $H_3$, we introduce the new variables $x'=x-B_\eps(t)$, $y'=y-B_\eps(t)$, and let
\begin{equation*}
E(x',y') = \frac 12 \theta_\eps(x'+B_\eps(t),t)\, \theta_\eps(y'+B_\eps(t),t) \, [\nabla W_h(x')-\nabla W_h(y')] \cdot K(x'-y') \;,
\end{equation*} 
so that $H_3 = \int\!\rmd x' \! \int\!\rmd y'\,E(x',y')$. We observe that $E(x',y')$ is a symmetric function of $x'$ and $y'$ and that, by \eqref{W1}, a necessary condition to be different from zero is if either $|x'|\ge h$ or $|y'|\ge h$. Therefore, 
\begin{equation*}
\begin{split}
H_3  &= \bigg[ \int_{|x'| > h}\!\rmd x' \! \int\!\rmd y' + \int\!\rmd x' \! \int_{|y'| > h}\!\rmd y' -  \int_{|x'| > h}\!\rmd x' \! \int_{|y'| > h}\!\rmd y'\bigg]E(x',y') \\ & = 2 \int_{|x'| > h}\!\rmd x' \! \int\!\rmd y'\,E(x',y')  -  \int_{|x'| > h}\!\rmd x' \! \int_{|y'| > h}\!\rmd y'\,E(x',y') \\ & = H_3' + H_3'' + H_3'''\;.
\end{split}
\end{equation*}
with 
\begin{equation*}
\begin{split}
H_3' & = 2 \int_{|x'| > h}\!\rmd x' \! \int_{|y'| \le h/2}\!\rmd y'\,E(x',y') \;, \\
  H_3'' & = 2 \int_{|x'| > h}\!\rmd x' \! \int_{|y'| > h/2}\!\rmd y'\,E(x',y')\;, \\ H_3''' & = -  \int_{|x'| > h}\!\rmd x' \! \int_{|y'| > h}\!\rmd y'\,E(x',y')\;.
\end{split}
\end{equation*}
By the assumptions on $W_h$, we have $\nabla W_h(z) = \eta_h(|z|) z/|z|$ with $\eta_h(|z|) =0$ for $|z| \le h$. In particular, $\nabla W_h(y') = 0$ for $|y'| \le h/2$. Therefore, 
\begin{equation*}
H_3' =  \int_{|x'| > h}\!\rmd x' \, \theta_\eps(x'+B_\eps(t),t) \eta_h(|x'|) \,\frac{x'}{|x'|}\cdot  \int_{|y'| \le h/2}\!\rmd y'\, K(x'-y') \, \theta_\eps(y'+B_\eps(t),t)\;.
\end{equation*}
In view of  \eqref{W2}, $|\eta_h(|z|)| \le C_1/h$, so that 
\begin{equation*}
|H_3'| \le \frac{C_1}{h} m_t(h) \sup_{|x'| > h} \bigg|\frac{x'}{|x'|}\cdot  \int_{|y'| \le h/2}\!\rmd y'\, K(x'-y') \, \theta_\eps(y'+B_\eps(t),t)\bigg| \;.
\end{equation*}
We now observe that the expression inside the modulus in the right-hand side is equal to the term $H_1$ in \eqref{acca_uno} (with $h$ in place of $R_t$), which has been bounded in \eqref{acca_uno'}-\eqref{acca_uno''} (it is readily seen that the proof works also if the assumption  $|x'|=R_t$ is relaxed to $|x'|\ge R_t$). We conclude that
\begin{equation}
|H_3'| \le \frac{C\, I_\eps(t)}{ h^{4+\alpha}} \, m_t(h)\;.
\label{h3'}
\end{equation}
Now, by \eqref{W3},
\begin{equation*}
\begin{split}
|H_3''| + |H_3'''| & \le \frac{C}{ h^2} \int_{|x'| \ge h}\!\rmd x' \, \theta_\eps(x'+B_\eps(t),t)  \! \int_{|y'| \ge h/2}\!\rmd y'\,\frac{\theta_\eps(y'+B_\eps(t),t)}{|x'-y'|^\alpha}  \\ & \leq  \frac{C}{h^2} \, m_t(h) \, M \pi \eta^{(2-\alpha)} \epsilon^{-2}\;,
\end{split}
\end{equation*}
where in the last integral we have rearranged again the  active scalar's mass as close as possible to the singularity $y'=x'$, 
and we have integrated over a circle of radius $\eta$ around the singularity.
We have used also the initial hypothesis $|\theta_\epsilon(x,0)|\leq M \epsilon^{-2}$ and the fact that $\theta$
is conserved along the motion.
The radius $\eta$ of the circle is obtained by the constraint $m_t(h/2)= M\pi \eta^2 \epsilon^{-2}$.
We use then the bound 
$$
m_t(h/2) \leq C \frac{I_\epsilon(t)}{h^2}
$$
to obtain
$$
\eta \leq C \,  \frac{\sqrt{\epsilon^2 \, I_\epsilon(t)}}{h}\;.
$$
Therefore,
\begin{equation}
\label{h3s}
|H_3''|+|H_3'''| \le  \frac{C}{h^{4-\alpha}} m_t(h) \epsilon^{-\alpha }  {I_\epsilon(t)}^{(1-\frac{\alpha}{2})}\;.
\end{equation}

Concerning $H_4$, we observe that by \eqref{W1} the integrand is different from zero only if $h\le |x-B_\eps(t)|\le 2h$. Therefore, by \eqref{Lip_const} and \eqref{W2} we have, 
\begin{equation*}
\begin{split}
|H_4| & \le \frac{C_3}{h} 2\|F\|_\infty \int_{|x'|\ge h}\!\rmd x' \theta_\eps(x'+B_\eps(t),t) \int_{|y'|\ge h}\!\rmd y'\,\theta(y'+B_\eps(t),t) \\ & \quad + \frac{C_3}{h} D_t \int_{h \le |x'|\le 2h}\!\rmd x' \theta_\eps(x'+B_\eps(t),t)\int_{|y'| \le h}\!\rmd y'\,|x'- y'| \, \theta(y'+B_\eps(t),t)\;.
\end{split}
\end{equation*}
Since $|x'-y'| \le 3h$ in the domain on integration of the last integral and using the Chebyshev's inequality in the first one we get,
\begin{equation}
\label{h4s}
|H_4| \le \frac{2C_3 \|F\|_\infty I_\eps(t)}{h^3} m_t(h) + 3C_3 D_t m_t(h)\;.
\end{equation}

By \eqref{mass 4}, \eqref{h3'},  \eqref{h3s}, \eqref{h4s}, and using that $D_t\le D$, see \eqref{Lip_const}, we have,
\begin{equation}
\label{2mass 4''}
\frac{\rmd}{\rmd t} \mu_t(h) \le A(h) m_t(h) \quad\forall\, t \in [0,\gamma |\log\eps|]\;,
\end{equation}
where, for some $C_4>0$,
\begin{equation}
\label{mass 4bis}
A(h) = C_4 \bigg(\frac{\eps^\delta}{h^{4+\alpha}}+ \frac{\epsilon^{\delta(1-\frac{\alpha}{2})-\alpha}}{h^{4-\alpha}} +\frac{\eps^\delta}{h^3} + 1\bigg)\;,
\end{equation}
having used \eqref{I_eps} to bound $I_\epsilon(t)$.  The parameter $\gamma$ is chosen suitably small, in order to have $\delta=2-2D\gamma$ close to $2$
(which will be used hereafter).

Given 
\begin{equation}
0<\beta< \bar\beta=\frac{\delta\left(1-\frac{\alpha}{2}\right)-\alpha}{4-\alpha} <\frac12
\label{uno*}
\end{equation}
with $\gamma$ so small that
$$
\delta\left(1-\frac{\alpha}{2}\right)-\alpha>0\;, \qquad   \textnormal{that is}   \qquad    \delta>\frac{\alpha}{1-\frac{\alpha}{2}}\;,
$$
we fix 
\begin{equation}
\beta_* \in \left(\beta,\bar\beta \right).
\label{due*}
\end{equation}
Hence, for $h\geq \epsilon^{\beta_*}$, it results   $A(h)\le A_*$, for a suitable constant
$A_*>0$. Therefore, by \eqref{2mass 3} and \eqref{2mass 4''},
\begin{equation}
\label{mass 14'}
\mu_t(h) \le \mu_0(h) + A_* \int_{0}^t \rmd s\, \mu_s(h/2) \quad \forall\,  t \in [0,\gamma|\log\eps|]\quad \forall\, h\ge \eps^{\beta_*}\;,
\end{equation}
which can be iterated $n$ times, provided $2^{-n}h\ge \eps^{\beta_*}$,  so that, for any $t\in [0,\gamma|\log\eps|]$, 
\begin{equation}
\label{mass 15'}
\begin{split}
\mu_t(h) & \le \mu_0(h) + \sum_{j=1}^n \mu_0(2^{-j}h) \frac{(A_*t)^j}{j!} + \frac{A_*^{n+1}}{n!} \int_0^t\!\rmd s\,  (t-s)^n\mu_s(2^{-(n+1)}h) \\ & = \frac{A_*^{n+1}}{n!} \int_0^t\!\rmd s\,  (t-s)^n\mu_s(2^{-(n+1)}h) \le  \frac{(A_*\gamma |\log\eps |)^{n+1}}{(n+1)!}\;,
\end{split}
\end{equation}
where we used that since $\Lambda_\eps(0) \subset \Sigma(z_*, \eps)$ and $\eps <1$ then $\mu_0(2^{-j}h)=0$ for any $j=0,\ldots,n$, and that $\mu_s(2^{-(n+1)}h)\le 1$. By applying \eqref{mass 15'} with $h=\eps^\beta$, $n=\lfloor (\beta_*-\beta)|\log_2\eps| \rfloor$,  and using the Stirling approximation for $(n+1)!$, we obtain that, given any $\ell>0$, for $\gamma$ small enough, 
\begin{equation*}
\lim_{\eps\to 0} \eps^{-\ell} \mu_t(\eps^\beta) = 0\quad \forall\, t \in [0,\gamma|\log\eps|]\;,
\end{equation*}
which concludes the proof. 
$\square$

\bigskip

{\textit{Proof of Theorem \ref{thm:2}}}.

By \eqref{I_eps}, \eqref{R_ti}, \eqref{eq_maxspost}, and recalling $D_t\le D$, see \eqref{Lip_const}, we have, whenever $|x(x_0,t)-B_\eps(t)| = R_t$,
\begin{equation}
\label{stimr}
\frac{\rmd}{\rmd t} |x(x_0,t)- B_\eps(t)| \le 
2 D R_t +\frac{4 C_1 \, \epsilon^\delta}{R_t^{3+\alpha}}+
\frac{C_2}{1-\alpha}\left[\left(M\epsilon^{-2}\right)^{\frac{1+\alpha}{1-\alpha}}\, m_t(R_t/2)\right]^{\frac{1-\alpha}{2}}
\end{equation}
for all  $t\in [0,\gamma|\log\eps|]$,
 $\delta=2-2D\gamma$ and $\gamma>0$ small. This implies that $\Lambda_\eps(t) \subset \Sigma(B_\eps(t),  {\mathcal{R}}(t))$ for any $t\in [0,\gamma|\log\eps|]$, where ${\mathcal{R}}(t)$ is a solution to 
\begin{equation}
\label{stimrbis}
\dot  {\mathcal{R}}(t) = 2 D {\mathcal{R}}(t) +\frac{4 C_1 \,  \epsilon^\delta}{{\mathcal{R}}(t)^{3+\alpha}}+
\frac{C_2}{1-\alpha}\left[\left(M\epsilon^{-2}\right)^{\frac{1+\alpha}{1-\alpha}}\, m_t({\mathcal{R}}(t)/2)\right]^{\frac{1-\alpha}{2}} \;.
\end{equation}
Hence we have $R_t\leq {\mathcal{R}}(t)$ for any $t\in[0, \gamma|\log\eps|]$.
For ${\mathcal{R}}(t)\geq \epsilon^{\beta'}$, $\beta'\in(0, \bar\beta)$, the last two terms on the right hand side of \eqref{stimrbis} are negligible with respect to ${\mathcal{R}}(t)$,
by Lemma \ref{lem:3} and by the following
$$
\frac{4 C_1 \,  \epsilon^\delta}{{\mathcal{R}}(t)^{3+\alpha}} \leq \frac{4 C_1 \,  \epsilon^\delta}{\epsilon^{(3+\alpha)\beta'}}
= 4 C_1 \, \epsilon^{\delta-(3+\alpha)\beta'}\leq \epsilon^{\beta'}
$$
which holds true for $\epsilon$ sufficiently small (the estimate is analogous to that of the first term in the r.h.s. of \eqref{mass 4bis}).

Hence, for ${\mathcal{R}}(t) \geq \epsilon^{\beta'}$, after a certain time $t_0$
(of course if such $t_0$ does not exist the Theorem is already proved),
$$
\dot {\mathcal{R}}(t) \leq C_5\, {\mathcal{R}}(t)  \Rightarrow{\mathcal{R}}(t) \leq \epsilon^{\beta'}  {\rm{e}}^{C_5(t-t_0)}
$$
and, for $t-t_0\leq \gamma |\log\eps|$
$$
{\mathcal{R}}(t) \leq \epsilon^{\beta'} {\rm{e}}^{-C_5 \gamma \log\eps} = \epsilon^{\beta' - C_5\gamma} = \epsilon^{\beta}
$$
with $\beta = \beta' - C_5\gamma$ and $0<\gamma < \beta'/C_5$. Note that (see \eqref{r_boun})
$$
\bar\beta < \frac{2-2\alpha}{4-\alpha}
$$
since $\delta$ (in the definition of $\bar\beta$ \eqref{barbeta}) is less than $2$.
This concludes the proof of Theorem \ref{thm:2}, with $\zeta_1=\gamma$.
$\square$

\bigskip

The proof of Theorem \ref{thm:1} follows quite immediately from Theorem \ref{thm:2},
and we address to \cite{BuM18} for the details.

\section{Improvements }

\subsection{Power law }

In the preview Sections we have obtained a logarithmic law for the persistence of the localization. For some particular initial data we can improve this result. We give here two examples.

\noindent i) A single patch. 

\noindent   Consider a single blob of active scalar in $\bb R^2$, with compact support and alone in the plane. As  time goes by  the support could increase (see \cite{Garra} for a detailed analysis about this point). 
The key point is that the blob is not subject to an external field ($F(x,t)$ is absent, and then its Lipschitz constant $D_t$ is also zero).
In this case we can obtain a power-law lower bound on the maximal time for which
the localization persists, in complete analogy to the Euler case ($\alpha=0$).

\noindent ii) Pseudo-vortices going to infinity.

\noindent Consider initial data localized around $N$ points $z_i(0)$, which move according to the differential equations
 $$
 a_i\dot{z}_{i_{1}} = \frac{ \partial {H}}{\partial {z}_{i_{2}}} \;, \qquad  a_i\dot{z}_{i_{2}} = -\frac{ \partial {H}}{\partial {z}_{i_{1}}} \;,
 $$
 where   $z_i=({z}_{i_{1}},  {z}_{i_{2}} )$  and
 $$
 H=- \frac{1}{2\pi}\sum_{i,j;i\ne j}^N a_i a_j |z_i-z_j|^{-\alpha} \; , \quad 0<\alpha < 1 \;.
 $$

Some properties of this dynamical system are given in Section 3 of \cite{CGM13}
and in \cite{ash}. Moreover, there are particular initial data $z_i(0)$, values of $N$, and values of $a_i$, for which $|z_i(t)-z_j(t)|$ go to infinity as $t \to \infty$. An explicit example has been given in Section 4 of \cite{LuS19}. In this case case we can prove a power law for the time scale in which the localization property holds. In fact it results (see \cite{LuS19}, Section 4) that
for $N=3$, there are particular initial data $z_i(0)$ and intensities $a_i$ for which
the three pseudo-vortices,
initially posed on the vertices of a triangle of sides of length $L_{ij}$, for successive times remain posed in the vertices of a triangle of sides of length $L_{ij}(t)$, where
\begin{equation}
\label{3point}
L_{ij}(t) = L_{ij}(0) {\left(1+gt\right)}^\frac{1}{2+\alpha}\;, \quad g>0\;, 
\end{equation}
that is, the triangle grows in the future (and shows a collapse for $t=-g^{-1}$), but remains similar in form.  This behavior is analogous to what happens in the well known Euler case ($\alpha=0$, see  \cite{Aref} or, for instance,  \cite{MaP94}).

The field $F(x,t)$ under which one pseudo-vortex moves is generated by the other two, and it satisfies,
for suitable constants $b,L>0$,
\begin{equation}
\label{bounded}
|F(x,t)| \le \frac{b}{L_{ij}(t)^{\alpha+1}}\;,
\end{equation}
\begin{equation}
\label{Lips}
|F(x,t)-F(y,t)| \le D_t |x-y|\;, 
\end{equation}
where the Lipschitz constant $D_t$ is such that
\begin{equation}
D_t\le\frac{L}{L_{ij}(t)^{\alpha+2}}=  \frac{L}{\left[  L_{ij}(0) \left({1+gt}\right)^\frac{1}{2+\alpha}   \right]^{\alpha+2}} =  \frac{const}{1+gt} \;.
\end{equation}
The decreasing in time of the Lipschitz constant $D_t$ does not depend on $\alpha$,
and it is the same as for the Euler case $\alpha=0$. The improvement of the content of Theorems \ref{thm:1} and \ref{thm:2} is closely related to the decreasing property of the Lipschitz constant,
so it is the same as analysed in \cite{BuM18} (to which we address for the proof),
leading to  $T_{\epsilon, \beta}>\epsilon^{-\zeta_0}$,    $\forall \epsilon\in(0, \epsilon_0)$,
for suitable $\epsilon_0>0$ and $\zeta_0>0$.

\subsection{Vanishing viscosity limit}

It is possible to introduce the so-called modified  dissipative  surface quasi-geostrophic equation,
\begin{equation}
\partial_t\theta +u\cdot \nabla \theta + k(-\Delta)^{\gamma/2}(\theta) = 0 \; , \qquad k>0 \; ,
\label{viscous}
\end{equation}
(see for instance \cite {Const2, May, R19, Wu97}) and study the limit $k \to 0$ as $\epsilon \to 0$ in the initial condition.

\noindent When $\gamma = 2$ and $\alpha =0$ we obtain the vanishing viscosity limit for the Navier-Stokes equation for singular initial data, that has been discussed in general in 
\cite{Mar90,Mar98,Mar07,Gal11} and in a situation similar to that of the present paper in \cite {CeS18},
while for $\alpha\in [0,1)$ and $\gamma\in [0,2]$ the localization of solutions to  \eqref{viscous}
is discussed in \cite {R19}.
Of course in presence of viscosity the localization happens when the vorticity out of a small region is not zero (as in the Euler case) but only very small.

\noindent It would be  reasonable to extend the analysis of the persistence
for long times of the localization when $\alpha>0$.

\bigskip
\noindent {\textbf{Acknowledgments}}.
Work performed under the auspices of GNFM-INDAM and the Italian Ministry
of the University (MIUR).

\bigskip
\bigskip
\begin{minipage}[t]{10cm}
\begin{flushleft}
\small{
\textsc{Guido Cavallaro}
\\*Sapienza Universit\`a di Roma,
\\*Dipartimento di Matematica
\\*Piazzale Aldo Moro, 2
\\* Roma, 00185, Italia
\\*e-mail: cavallar@mat.uniroma1.it
\\[0.4cm]
\textsc{Roberto Garra}
\\*Sapienza Universit\`a di Roma,
\\*Dipartimento di Scienze Statistiche
\\*Piazzale Aldo Moro, 2
\\* Roma, 00185, Italia
\\*e-mail: roberto.garra@sbai.uniroma1.it
\\[0.4cm]
\textsc{Carlo Marchioro}
\\*International Research Center M\&MOCS,
\\*Universit\`a di L'Aquila
\\*Palazzo Caetani
\\* Cisterna di Latina (LT), 04012, Italia
\\*e-mail: marchior@mat.uniroma1.it
}
\end{flushleft}
\end{minipage}


\end{document}